\documentclass[aps,twocolumn,PRL,reprint,superscriptaddress]{revtex4}
\usepackage{amsmath}
\usepackage[percent]{overpic}
\usepackage{times}
\usepackage{subfigure}
\usepackage{latexsym}
\usepackage{graphicx}
\usepackage{bm}
\usepackage{amssymb}
\usepackage{hyperref}
\begin{document}
\title{Hopf-link multi-Weyl-loop topological semimetals }
\author{Yao Zhou}
\affiliation{National Laboratory of Solid State Microstructures, Department of Physics, Nanjing University, Nanjing 210093, China}
\author{Feng Xiong}
\affiliation{National Laboratory of Solid State Microstructures, Department of Physics, Nanjing University, Nanjing 210093, China}
\author{Xiangang Wan}
\affiliation{National Laboratory of Solid State Microstructures, Department of Physics, Nanjing University, Nanjing 210093, China}
\affiliation{Collaborative Innovation Center of Advanced Microstructures, Nanjing University, Nanjing 210093, China}
\author{Jin An}
\email{anjin@nju.edu.cn}
\affiliation{National Laboratory of Solid State Microstructures, Department of Physics, Nanjing University, Nanjing 210093, China}
\affiliation{Collaborative Innovation Center of Advanced Microstructures, Nanjing University, Nanjing 210093, China}
\begin{abstract}
  We construct a generic two-band model which can describe topological Weyl semimetals with multiple closed Weyl loops. All the existing multi-Weyl-loop semimetals including the nodal-net, or nodal-chain and Hopf-link states can be examined within one same framework. Based on a two-loop model, the corresponding drumhead surface states for these topologically different bulk states are studied and compared with each other. The connection of our model with Hopf insulators is also discussed. Furthermore, to identify experimentally these topologically different Weyl semimetal states, especially distinguish the Hopf-link from unlinked ones, we also investigate their Landau levels. It is found that the Hopf-link state can be characterized by the existence of a quadruply degenerate zero-energy Landau band, regardless of the direction of the magnetic field.
\end{abstract}

\date{\today}

\maketitle

\emph{Introduction.}---As topological materials\cite{Hasan,qi2011topological,bansil2016colloquium,chiu2016classification}, gapless topological semimetals, especially Weyl semimetals have recently attracted widespread attentions. According to the dimensionality of the manifolds of crossings between the conduction band and valence band, Weyl semimetals can be classified into Weyl-point\cite{wan2011topological,hosur2013recent,weng2015weyl,lv2015experimental,huang2015weyl,shekhar2015extremely,yang2015weyl,xu2015discovery,lu2015experimental,jia2016weyl} and Weyl-loop semimetals\cite{burkov2011topological,carter2012semimetal,phillips2014tunable,chiu2014classification,weng2015topological,fang2015topological,chen2015topological,chen2015nanostructured,xie2015new,yamakage2015line,chan20163,zhao2016topological,ezawa2016loop,bian2016topological,xu2017topological,quan2017single}. Not only has the concept of the former now been extended to type II \cite{soluyanov2015type,wang2016mote,autes2016robust,koepernik2016tairte,belopolski2016fermi,chan2016type,chang2016strongly}, but also that of the latter has been promoted to the nodal-net\cite{zeng2015topological,kim2015dirac,yu2015topological,feng2017topological,kobayashi2017crossing} or nodal-chain\cite{bzduvsek2016nodal,yu2017nodal} cases. Very recently, a new family member, namely, the Hopf-link Weyl semimetal\cite{chen2017topological,yan2017nodal,ezawa2017topological,chang2017weyl,sun2017double,chang2017topological,bi2017nodal} has been found, where the two bands touch each other at two closed loops which form a Hopf link. While the Weyl-point topological semimetals have Fermi-arc surface states\cite{wan2011topological,xu2015discovery1}, the Weyl-loop topological semimetals have robust drumhead surface states \cite{burkov2011topological,chan20163,matsuura2013protected,bian2016drumhead}, which enables the possibility of surface high-temperature superconductivity\cite{kopnin2011high}.

In previous studies, all these Weyl-loop topological semimetals are described by independent models, and we still lack an universal description and understanding of all these topologically different states within one same framework. In this paper, we newly construct a generic multi-Weyl-loop model which is capable of describing all the existing Weyl-loop topological semimetals, including the nodal-net, nodal-chain and Hopf-link Weyl semimetals. Within this generic model, we examine and identify these topologically distinct bulk states and their corresponding drumhead surface states protected by the chiral symmetry. A simple model for a three-loop Weyl semimetal is illustrated with any two of the loops linked with nontrivial linking numbers. We also found that each Hopf-link model can be connected with a nontrivial Hopf insulator. Furthermore, it is shown that these topologically different Weyl-loop states can be distinguished by their Landau levels or by the nodes of their corresponding gapless Floquet states driven by a circularly polarized light.

\emph{Hamiltonian model.}---We propose a new two-band model for Weyl semimetals with multiple closed Weyl loops. The general model Hamiltonian $\mathcal{H}_{n}(\bm{k})$ with $n$ Weyl loops can be written as,
\begin{eqnarray}\label{Hn}
  \mathcal{H}_{n}(\bm{k})&=&\left[
	\begin{array}{cc}
	  0 & q(\bm{k}) \\
	  q^{*}(\bm{k})& 0 \\
	\end{array}
  \right],
\end{eqnarray}
where $q(\bm{k})=\prod\limits_{i=1}^{n} \mathcal{F}_{i}(\bm{k})$. For $i=1,2,...n$, each $\mathcal{F}_{i}(\bm{k})$ is a complex function of $\bm{k}$ and each equation $\mathcal{F}_{i}(\bm{k})=0$ is assumed to determine a closed loop. The energy dispersion of $\mathcal{H}_{n}$ reads as:

\begin{equation}
  \begin{split}
	\mathcal{E}(\bm{k})&=\pm \vert q (\bm{k}) \vert=\pm \prod\limits_{i=1}^{n}\vert\mathcal{F}_{i}(\bm{k})\vert.
  \end{split}
\end{equation}
Obviously, the zero-energy states are composed of the N closed loops determined by $\mathcal{F}_{i}(\bm{k})=0$ with $i=1,2,...n$. The advantage of this $N$-loop model is that since each loop can be constructed independently, any two loops can be linked to each other with nontrivial linking numbers, or all loops can be connected to form a chain or net. Therefore by utilizing our two-band model, one can study the nodal-chain, nodal-net and Hopf-link semimetals within one same framework.

\begin{figure*}[ht]
  \begin{center}
	\includegraphics[width=18cm,height=12cm]{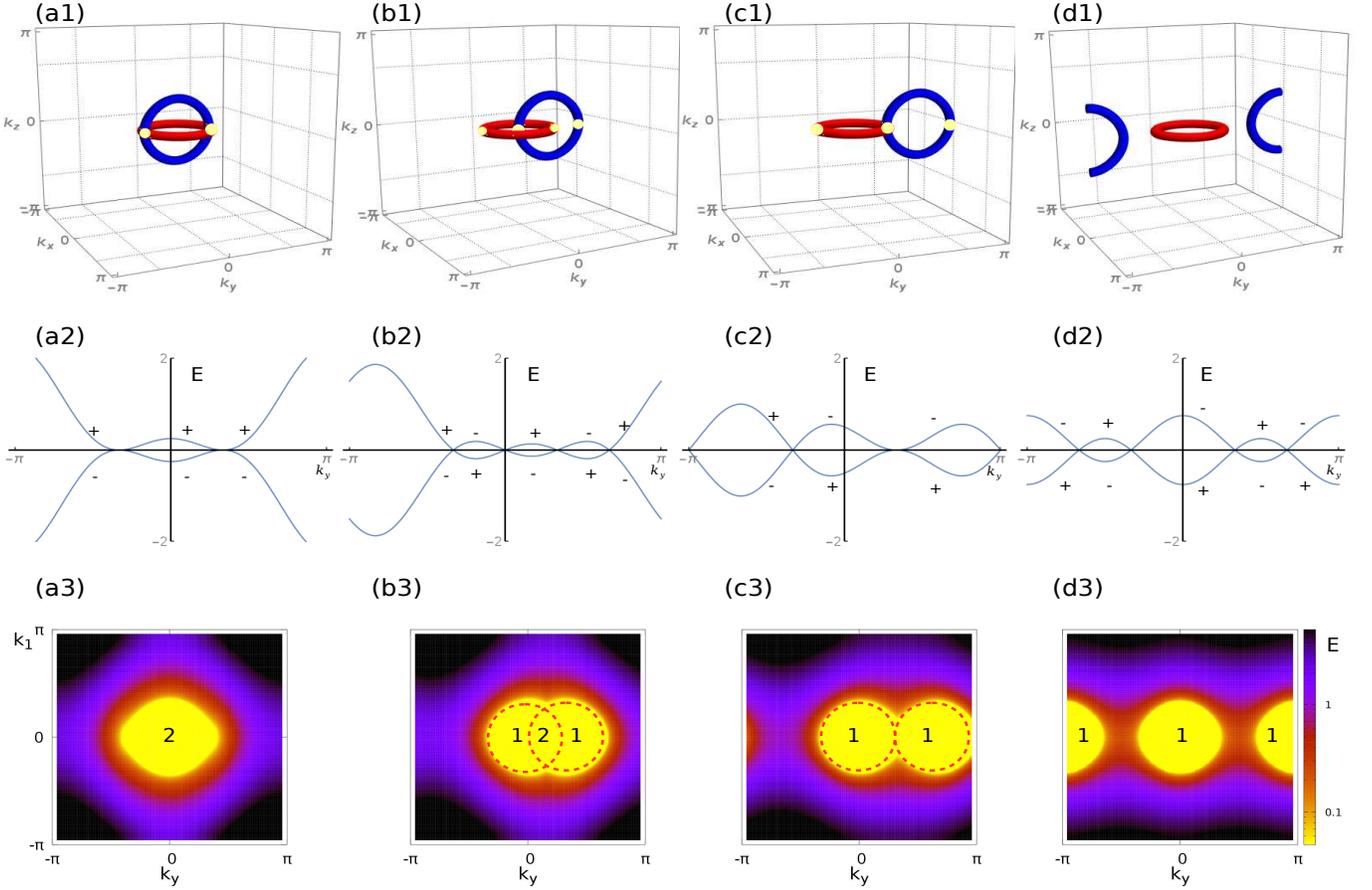}	
  \end{center}
  \vspace{-0.4cm}
  \caption{(Color online)Two-loop Weyl semimetals with their structures adjusted by parameter $k_{y}^{0}$. (a1)The nodal-net state with $k_{y}^{0}=0$. (b1)The Hopf-link state with $k_{y}^{0}=\frac{\pi}{3}$. (c1)The nodal-chain state with  $k_{y}^{0}=\frac{2\pi}{3}$. (d1)The unlinked state with $k_{y}^{0}=\pi$. (a2)-(d2)The corresponding bulk energy bands along $k_{y}$ axis, where $\pm$ denote the signs of the eigenvalues of rotation operator $\mathcal{C}_{2}$ for the bands. (a3)-(d3)The corresponding drumhead surface states for boundary along direction $(1,0,-1)$. The integers are the 1D winding numbers $N_{\mathcal{L}}$ with path $\mathcal{L}$ chosen along $(1,0,-1)$, with their absolute values representing degeneracies of boundary states. The colored dots in (a1)-(c1) are the nodes of the corresponding Floquet gapless states driven by a circularly polarized light propagating along $y$ direction. Here $m=2.5$.
  } \label{fig1}
\end{figure*}

\emph{Symmetry analysis}.--- The generic two-band Hamiltonian $\mathcal{H}_{n}$ has the combined $\mathcal{PT}$ symmetry \cite{zhao2016unified}, $\mathcal{CP}$ symmetry and chiral symmetry respectively,

\begin{equation}
  (\sigma_{x}\mathcal{K})\mathcal{H}_{n}(\bm{k})(\sigma_{x}\mathcal{K})^{-1}=\mathcal{H}_{n}(\bm{k}),
\end{equation}
\begin{equation}
  (\sigma_{y}\mathcal{K})\mathcal{H}_{n}(\bm{k})(\sigma_{y}\mathcal{K})^{-1}=-\mathcal{H}_{n}(\bm{k}),
\end{equation}
\begin{equation}
  \{\mathcal{H}_{n}(\bm{k}),\sigma_{z}\}=0,
\end{equation}
where $\mathcal{K}$ is complex conjugation. Consider a closed path $\mathcal{L}$ within the gapful region in the BZ. Due to the chiral symmetry, the Hamiltonian $\mathcal{H}_{n}$ restricted to the path is a $1D$ system which belongs to symmetry class \textbf{A}\uppercase\expandafter{\romannumeral3} \cite{schnyder2008classification,ryu2010topological,schnyder2011topological}, resulting in a well defined 1\textbf{D} winding number,
\begin{equation}
  N_{\mathcal{L}} = \frac{1}{2\pi i} \oint_{\mathcal{L}} dl Tr[q^{-1} (\bm{k}) \bigtriangledown_{l} q(\bm{k}) ].
\end{equation}
Depending on the path chosen which encloses a few of the Weyl loops or not, $N_{\mathcal{L}}$ can be nontrivial or trivial.

\emph{Hopf-link Weyl semimetal without mirror symmetry}.---Now let's consider the following minimal two-loop Hamiltonian:
\begin{eqnarray}\label{H2}
  \mathcal{H}_{2}(\bm{k})&=&\left[
	\begin{array}{cc}
	  0 & \mathcal{F}_{1}(\bm{k})\mathcal{F}_{2}(\bm{k}) \\
	  \mathcal{F}^{*}_{1}(\bm{k})\mathcal{F}^{*}_{2}(\bm{k})& 0 \\
	\end{array}
  \right],
\end{eqnarray}
where $\mathcal{F}_{1}(\bm{k})$, $\mathcal{F}_{2}(\bm{k})$ are chosen as,
\begin{equation}
  \mathcal{F}_{1}(\bm{k}) =\cos k_{x}+\cos k_{y}+\cos k_{z}-m + \bm{i}\sin k_{z},
\end{equation}
\begin{equation}
  \mathcal{F}_{2}(\bm{k}) =\cos k_{x}+\cos(k_{y}-k^{0}_{y})+\cos k_{z}-m+\bm{i}\sin k_{x}.
\end{equation}
$\mathcal{H}_{2}(\bm{k})$ owns two loops within plane $k_{z}=0$ and $k_{x}=0$ respectively, as shown in Fig.\ref{fig1}. Here $m$ and $k^{0}_{y}$ are adjustable parameters, with the first responsible for the size and the second for the relative position of the two loops. This minimal model is constructed here mainly to capture the essential physics of the two-loop Weyl semimetals. To relate the two-loop model to real materials, one may seek more complicated $\mathcal{F}_{1}(\bm{k})$, $\mathcal{F}_{2}(\bm{k})$ so that the Hamiltonian $\mathcal{H}_{2}(\bm{k})$ has proper crystal symmetry.

When varying parameter $k^{0}_{y}$ continuously, the system becomes successively the nodal-net, Hopf-link, nodal-chain and unlinked topological semimetals. Different from the four-band model with mirror symmetry in Ref\cite{chang2017topological}, this two-loop model is two-band without mirror symmetry. However, $\mathcal{H}_{2}$ has a $\mathcal{C}_{2}$ rotation symmetry with respect to $k_{y}$ axis,
\begin{equation}
  \mathcal{C}_{2}^{\dagger}\mathcal{H}_{2}(k_{x},k_{y},k_{z})\mathcal{C}_{2} =\mathcal{H}_{2}(-k_{x},k_{y},-k_{z}),
\end{equation}
where $\mathcal{C}_{2}=\sigma_{x}$. Thus along $k_{y}$ axis, we have the commutation relation $[\mathcal{H}_{2}(k_{x}=0,k_{y},k_{z}=0),\mathcal{C}_{2}]=0$, indicating the two bands of the 1D system as functions of $k_{y}$ can be labeled by the eigenvalues $\pm1$ of $\mathcal{C}_{2}$. This is exhibited in Fig.\ref{fig1}.(a2)-(d2), where for the upper band of the four topologically distinct states, the sequences of the eigenvalues' signs are $(+,+,+)$, $(+,-,+,-,+)$, $(+,-,-)$, $(-,+,-,+,-)$, respectively, as scanning $k_{y}$ from $-\pi$ to $\pi$. These sequences distinguish most of the topological semimetal states from others, with the exception of the Hopf-link state from the unlinked one.

\begin{figure}[ht]
  \begin{center}
	\includegraphics[width=8.5cm,height=6cm]{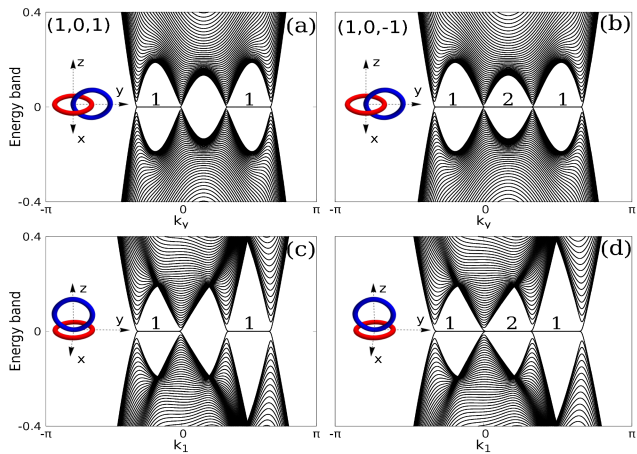}	
  \end{center}
  \vspace{-0.4cm}
  \caption{(Color online)Comparison of the surface bands for the Hopf-link Weyl seimimetal with that of the unlinked one. (a)(b)The surface bands for the Hopf-link state. (c)(d)The surface bands for the unlinked state, where $k_{1}=(k_{x}+k_{z})/2$. Open-boundary conditions are applied along $(1,0,1)$ for (a),(c), while along $(1,0,-1)$ for (b),(d). The insets show schematically the linking structures for the two bulk loops. The numbers labeled at the zero-energy flat band represent the 1D winding numbers.
  } \label{fig2}
\end{figure}

To explore the bulk-edge correspondence, we have also shown accordingly in Fig.\ref{fig1}.(a3)-(d3) the drumhead surface states, of which the profile is actually a combination of the 2D projections of the two bulk loops on the surface. These surface states are protected by the chiral symmetry, since within the drumhead area, each surface state corresponds to a nontrivial winding number $N_{\mathcal{L}}$ with path $\mathcal{L}$ chosen along the surface normal direction. Note that the existence of the surface states within the overlap region of the 2D projections for the two loops(as shown in Fig.\ref{fig1}.(b3)) is not the intrinsic property relevant to a Hopf link. To demonstrate this point, we apply open-boundary conditions along another direction. For the same bulk Hopf-link state, there would be no surface state within the overlap region. While for a unlinked state, there still exist surface states at the overlap regions for particular boundary directions, as shown in Fig.\ref{fig2}. Therefore it is hard to distinguish the Hopf-link semimetals from the unlinked ones merely by the existence or not of the surface states at the overlap region.

\begin{figure}[ht]
  \begin{center}
	\includegraphics[width=8.5cm,height=4cm]{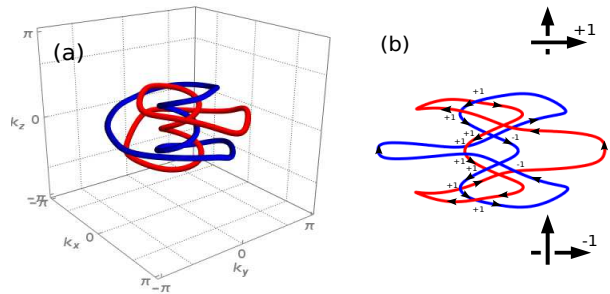}	
  \end{center}
  \vspace{-0.4cm}
  \caption{(Color online)(a)The linking structure of two loops with linking number 3. To calculate the linking number of this complex structure, a 2D projection of the two closed loops is exhibited in (b) with each crossing point labeled by $\pm1$. The linking number of the two loops is one half the absolute value of the sum over all these integers at crossing points.
  } \label{fig3}
\end{figure}

\begin{figure}[ht]
  \begin{center}
	\includegraphics[width=8.5cm,height=12cm]{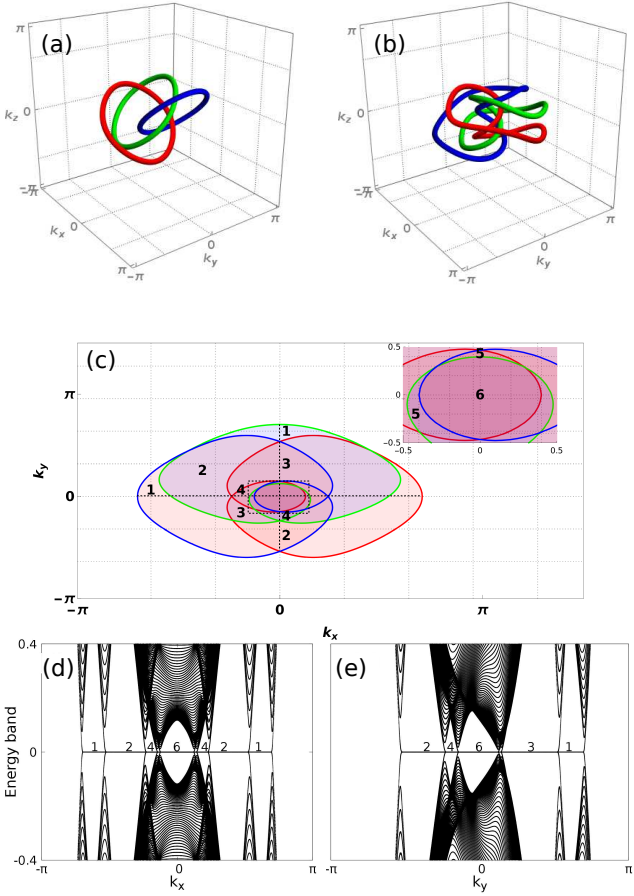}	
  \end{center}
  \vspace{-0.4cm}
  \caption{(Color online)Three nodal loops with any two of them linked with linking number 1 for (a) and 2 for (b). (c) The drumhead surface states of case (b) for boundary along $(0,0,1)$ direction, where the inset is the blowup of the central area. The corresponding surface bands scanning along $k_{x}$ axis and $k_{y}$ axis is exhibited in (d) and (e) respectively. The integers labeled in (c), (d) and (e) denote the 1D winding number $N_{\mathcal{L}}$ with path $\mathcal{L}$ chosen along $(0,0,1)$.} \label{fig4}
\end{figure}

\begin{figure}[ht]
  \begin{center}
	\includegraphics[width=8.5cm,height=8cm]{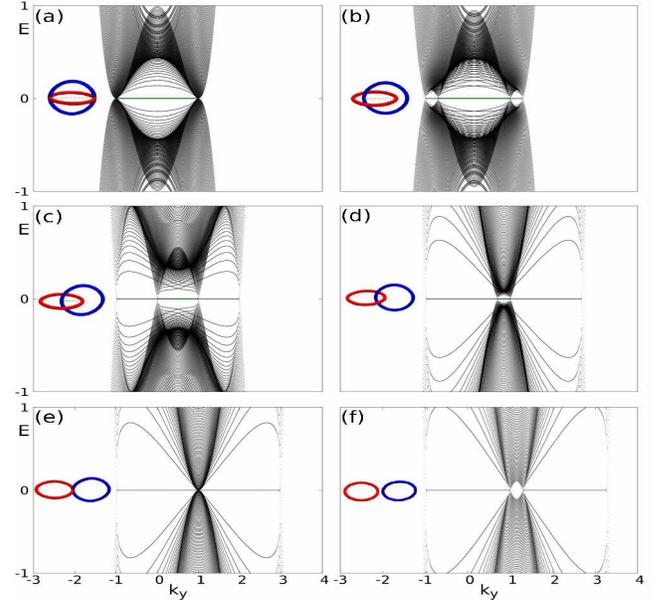}	
  \end{center}
  \vspace{-0.4cm}
  \caption{(Color online)The Landau spectra for the semimetals with two Weyl loops. (a)The nodal-net state with $k^{0}_{y}=0$. (b)-(d)The Hopf-link state with $k^{0}_{y}=0.3$, $1$, $ 1.7$ respectively. (e)The nodal-chain state with $k^{0}_{y}=2$. (f)The unlinked state with $k^{0}_{y}=2.3$. Here $k_{y}$ is measured in unit of the radius $k_{0}$ of the loops. The insets schematically show the topological configurations for the two loops, where the colored dotted lines indicate the $k_{y}$ regimes possessing a quadruply degenerate zero-energy landau level. Here $l_{B}k_{0}=14$.}
\label{fig5}
\end{figure}

\emph{Two Hopf-link loops with a high linking number }.---We can readily construct models for the two-loop Weyl semimetals, where the two Weyl loops are linked with a high linking number. $\mathcal{F}_{1}(\bm{k})$ can be chosen as follows,
\begin{equation}
  \mathcal{F}_{1}(\bm{k}) = (N_{1}+\bm{i}N_{2}) +(N_{3}+\bm{i}N_{4})^{q},
\end{equation}
where $q$ is an integer. $N_{1,2,3,4}$ are given as,
\begin{equation}
  \begin{split}
	N_{1}= \sin k_{x}, N_{2}= \sin k_{y}, N_{3}= \sin k_{z} \\
	N_{4} = m-\cos k_{x} -\cos k_{y} -\cos k_{z},
  \end{split}
\end{equation}
with $m$ the size parameter of the loops. These functions have been used to construct Hopf insulators or things related\cite{moore2008topological,deng2013hopf,deng2014systematic,wang2015z2,deng2018probe,kennedy2016topological,liu2017symmetry}. $\mathcal{F}_{2}(\bm{k})$ is obtained by rotating $\mathcal{F}_{1}(\bm{k})$ $180^{\circ}$ along $z$ axis. Without loss of generality, $m$ is chosen to be $2.3$. Generically, the linking number of the two closed curves is $q$. In Fig.\ref{fig3}(a) we exhibit the linking structure of the two loops with linking number $3$. When $q\geq3$, the linking structure is so complex that in order to determine the linking number, one has to make a 2D projection of the two closed loops. By associating each crossing point between the two loops' projections with an integer $+1$ or $-1$, mathematically, the linking number is one half the absolute value of the sum over all the integers, as shown in Fig.\ref{fig3}(b) for $q=3$.

\emph{Multiple Weyl loops linked with arbitrary linking numbers}.---From our Hamiltonian $(2)$, one can generically construct an $N$-loop model with arbitary linking numbers between any two of the loops. We take $N=3$ as an illustration. Here $\mathcal{F}_{1}(\bm{k})$ is chosen as before with $q=1$ or $q=2$. Then $\mathcal{F}_{2}(\bm{k})$, $\mathcal{F}_{3}(\bm{k})$ can be obtained by rotating $\mathcal{F}_{1}(\bm{k})$ along $z$ axis $90^{\circ}$ and $180^{\circ}$, respectively. The geometric configuration of three loops are shown in the Fig.\ref{fig4}(a)(b), where any two of them are linked with linking number $1$ and $2$ respectively. As before, the drumhead surface states are protected by the chiral symmetry and their profile is a combination of the 2D projections of the three closed loops, as shown in Fig.\ref{fig4}(c)-(e).

\emph{Connection with Hopf insulator}.---Now we connect our two-band model with a Hopf insulator\cite{wilczek1983linking,moore2008topological,deng2013hopf,deng2014systematic,wang2015z2,deng2018probe,kennedy2016topological,liu2017symmetry}. We still take the two-loop model $\mathcal{H}_{2}$ as an illustration. By adding an additional $\sigma_{z}$ term to $\mathcal{H}_{2}$, one can obtain the corresponding model for a Hopf insulator as follows:
\begin{equation}
  \mathcal{H}(\bm{k}) = \mathcal{H}_{2}(\bm{k}) + \frac{1}{2}(\vert\mathcal{F}_{1}(\bm{k})\vert^{2} - \vert\mathcal{F}_{2}(\bm{k})\vert^{2})\sigma_{z}.
\end{equation}
It can be easily seen that the energy band of $\mathcal{H}(\bm{k})$ is fully gapped as long as the two Weyl loops of $\mathcal{H}_{2}(\bm{k})$ share no crossing. The model can be expressed as $\mathcal{H}(\bm{k})=\bm{\sigma}\cdot(Re(\mathcal{F}_{1}\mathcal{F}_{2}),-Im(\mathcal{F}_{1}\mathcal{F}_{2}),\frac{1}{2}(\vert\mathcal{F}_{1}\vert^{2} - \vert\mathcal{F}_{2}\vert^{2}))$, which has the standard form for a Hopf insulator and thus defines a map from 3D BZ to a two-sphere, i.e., $T^{3}\rightarrow S^{2}$ \cite{moore2008topological,deng2013hopf,deng2014systematic}. Consider any two points on $S^{2}$, whose preimages would be two closed loops within BZ. The nontrivial Hopf insulator is characterized by an integer which is the linking number of the two loops. For the above model, we choose the north and south poles $(0,0,\pm1)$ on $S^{2}$. It can be readily checked that their preimages are respectively the two Weyl loops given by $\mathcal{F}_{1}=0$ and $\mathcal{F}_{2}=0$. Therefore if the two Weyl loops in the $T^{3}$ are linked(unlinked), the corresponding Hopf insulator is topologically nontrivial(trivial) with the linking number taken as its topological invariant.

\emph{Landau levels}.---To further identify the topologically different Weyl-loop semimetals, and especially to distinguish the Hopf-link state from the unlinked one, we study the Landau levels of the topological states shown in Fig.\ref{fig1}(a1)-(d1). To make things simple, we take the continuous limit of the model $\mathcal{H}_{2}(\bm{k})$. Thus $\mathcal{F}_{1}(\bm{k})$, $\mathcal{F}_{2}(\bm{k})$ become
\begin{equation}
\begin{aligned}
  \mathcal{F}_{1}(\bm{k}) =  1-k_{x}^{2}-k_{y}^{2}-k_{z}^{2} - \bm{i}k_{z}, \\
  \mathcal{F}_{2}(\bm{k}) =  1-k_{x}^{2}-(k_{y}-k^{0}_{y})^{2}-k_{z}^{2} - \bm{i}k_{x}.
\end{aligned}
\end{equation}
Here $\bm{k}$ and $k_{y}^{0}$ are measured in unit of the radius $k_{0}$ of the loops. The low-energy effective Hamiltonian under a magnetic field $\bm{B}=B\hat{e_{y}}$ is obtained by symmetrizing the momentum operators and then replacing $\bm{k}$ with $\bm{\Pi}=-i\nabla+e\bm{A}$\cite{yan2017nodal} in $\mathcal{H}_{2}(\bm{k})$, where Landau gauge $\bm{A}=(Bz,0,0)$ is assumed. Introducing the ladder operator $a=\frac{l_{B}}{\sqrt{2}\hbar}(\mathbf{\Pi}_{x} +i\mathbf{\Pi}_{z})$, with $l_{B}=\sqrt{\frac{\hbar}{eB}}$, the Landau levels as functions of the good quantum number $k_{y}$ can be numerically obtained and exhibited in Fig.\ref{fig5}.

The Landau spectrum is always symmetric with respect to $k_{y}=k^{0}_{y}$. There are two critical points of $k_{y}^{0}$ as displayed in Fig.\ref{fig5}(a) and (e) for the nodal-net and nodal-chain states, respectively. With a slight increase of $k^{0}_{y}$ for the nodal-net state with $k^{0}_{y}=0$, the two loops form a Hopf link and a quadruply degenerate zero-energy Landau band always exists. When the loop within plane $k_{x}=0$ is so shifted that the system actually becomes the nodal-chain state as shown in Fig.\ref{fig5}(e), the quadruply degenerate zero-energy Landau band disappears. The spectra for the Hopf-link loops are described by Fig.\ref{fig5}(b)-(d), while that for the unlinked by Fig.\ref{fig5}(f).

These numerical results can be understood as follows. The Landau levels for each $k_{y}=k$ are actually those of the 2D section of the original system cut by plane $k_{y}=k$. When there exists no crossing between the two loops and the plane, the 2D system is gapful which has no zero-energy Landau level. When the plane generically intersects one of the loops or both loops, the 2D system become gapless with $2$ or $4$ Dirac nodes respectively. Each pair of the Dirac nodes contributes a doubly degenerate zero-energy Landau level. Therefore, the Landau spectrum of the Hopf-link state is characterized by the existence of a quadruply degenerate zero-energy Landau band, since there always exist planes $k_{y}=k$ which intersect both loops, regardless of the direction of the magnetic field. On the other hand, for the unlinked state, the quadruply degenerate zero-energy Landau band does not exist for all the directions of the applied magnetic field. For some directions, as shown in Fig.\ref{fig5}(f), the unlinked state has only a doubly degenerate zero-energy Landau band. From this viewpoint, the two-loop nodal-net(nodal-chain) Weyl semimetals can be regarded as a critical state of the Hopf-link(unlinked) Weyl semimetals. With these distinct features, the Hopf-link states can also be distinguished clearly from the unlinked ones.


\emph{Floquet gapless states}.---Finally we investigate the low-energy effective Hamiltonian of the two-loop model studied above driven by a circularly polarized light\cite{oka2009photovoltaic,inoue2010photoinduced,lindner2011floquet,kitagawa2011transport}, which may help experimentalists identify the topologically distinct Weyl semimetals. Assume that the light is propagating along y direction which is parallel to both planes defined by the two loops, with $\bm{A}(t)=A_{0}(\cos\omega t,0,\sin\omega t)$. By the minimal coupling $\mathcal{H}(\bm{k})\rightarrow \mathcal{H}(\bm{k},t)=H(\bm{k}+e\bm{A}(t))$ , one can make the expansion $\mathcal{H}(\bm{k},t)=\sum_{n}\bm{H}_{n}(\bm{k})\exp^{\bm{i}n\omega t}$. Keeping to order of $\mathcal{O}(eA_{0})^{2}$, the effective Hamiltonian is derived as\cite{gu2011floquet,rudner2013anomalous,yan2016tunable,narayan2016tunable,taguchi2016photovoltaic,zhang2016theory},
\begin{equation}
\begin{aligned}
&\bm{H}_{\rm{eff}}=\bm{H}_{0}+\sum_{n\geq1}\frac{[\bm{H}_{n},\bm{H}_{-n}]}{n\omega}+\mathcal{O}(\frac{1}{\omega^{2}}) \\
&=[r_{1}r_{2}-k_{x}k_{z}+C^{2}(2k_{x}^{2}+2k_{z}^{2}-r_{1}-r_{2})]\sigma_{x}+ \\
&[r_{1}k_{x}+r_{2}k_{z}-2C^{2}(k_{x}+k_{z})]\sigma_{y}+\frac{\gamma\sigma_{z}}{\omega},
\end{aligned}
\end{equation}
where $C=eA_{0}$, and $\gamma=C^{2}[(k_{x}(2r_{2}^{2}+2r_{1}r_{2}-r_{1})-
  k_{z}(2r_{1}^{2}+2r_{1}r_{2}-r_{2})+2(k_{x}-k_{z})(k_{x}+k_{z})^{2}]$, with $r_{j}=Re\mathcal{F}_{j}(\bm{k})$, $j=1,2$. The effective Hamiltonian $\bm{H}_{\rm{eff}}$ describes a new gapless semimetal with several nodes located at the crossing points between the two loops and $k_{y}$ axis. For the states shown in Fig.\ref{fig1}(a)-(c), all the nodes(denoted by the colored dots) are located along $k_{y}$ axis, and the number of them is $2$, $4$ and $3$ respectively. This feature can thus be expected to be experimentally utilized to locate the positions of the Weyl loops.

\emph{Conclusion}---We have constructed a generic two-band model to describe Weyl semimetals with multiple Weyl loops. Since each loop can be constructed independently, by utilizing one same two-loop model, the drumhead surface states for topological Weyl semimetals ranging from the nodal-net, Hopf-link, nodal-chain and unlinked Weyl states have been studied and compared with each other. The topological semimetals described by the model can be connected with Hopf insulators and the model has also paved an easy way to explore Weyl semimetals with multiple Hopf-link loops. Experimentally, the fact that there always exists a quadruply degenerate zero-energy Landau band regardless of the direction of the applied magnetic field may help experimentalists to distinguish the Hopf-link states from unlinked ones.

\begin{acknowledgments}
We thank Feng Tang for useful discussions. This work was supported by NSFC under grants No.11174126, No.11525417, and the State Key Program for Basic Researches of China under No.2015CB921202.
\end{acknowledgments}

\bibliography{ref}

\end{document}